\documentclass[a4paper,11pt]{article}
\pdfoutput=1

\usepackage[utf8]{inputenc}
\usepackage[T1]{fontenc}
\usepackage[english]{babel}
\usepackage{layout}
\usepackage{setspace}
\usepackage{lmodern}
\usepackage{enumitem}
\usepackage{soul}
\usepackage{eurosym}

\usepackage{verbatim}
\usepackage{moreverb}

\usepackage{sectsty}

\usepackage{mathrsfs}
\usepackage{cancel}

\usepackage{makeidx}
\usepackage{multirow}
\usepackage{stmaryrd}

\usepackage{docstyle}

\usepackage{tikz,xcolor}

\definecolor{lime}{HTML}{A6CE39}
\DeclareRobustCommand{\orcidicon}{
	\begin{tikzpicture}
	\draw[lime, fill=lime] (0,0) 
	circle [radius=0.2] 
	node[white] {{\fontfamily{qag}\selectfont \tiny ID}};
	\draw[white, fill=white] (-0.0625,0.095) 
	circle [radius=0.007];
	\end{tikzpicture}
	\hspace{-2mm}
}

\foreach \x in {A, ..., Z}{\expandafter\xdef\csname orcid\x\endcsname{\noexpand\href{https://orcid.org/\csname orcidauthor\x\endcsname}
			{\noexpand\orcidicon}}
}

\renewcommand{\vec}[1]{\overrightarrow{\mathstrut #1}}
\newcommand{\erf}{\mathrm{erf}}
\newcommand{\erfc}{\mathrm{erfc}}
\newcommand{\sgn}{\mathrm{sgn}}

\newcommand{\be}{\begin{equation}}
\newcommand{\ee}{\end{equation}}
\newcommand{\bea}{\begin{eqnarray}}
\newcommand{\eea}{\end{eqnarray}}

\title{Fate of the false vacuum\\in string-inspired nonlocal field theory}

\author[a]{Anish Ghoshal}
\author[b]{and Florian Nortier}

\affiliation[a]{Institute of Theoretical Physics, Faculty of Physics, University of Warsaw,\\ul. Pasteura 5, 02-093 Warsaw, Poland}
\affiliation[b]{Université Paris-Saclay, CNRS, CEA, Institut de Physique Théorique,\\91191, Gif-sur-Yvette, France}

\emailAdd{anish.ghoshal@fuw.edu.pl}
\emailAdd{florian.nortier@ipht.fr}

\abstract{In this article, we study Coleman bounce in weakly nonlocal theories which are motivated from string field theory. The kinetic term is extended via an infinite series of high-order derivatives, which comes into play at an energy scale $M$, without introducing any new states or ghosts in the mass spectrum. We calculate the bubble nucleation in thin-wall approximation, treating the system in semi-classical manner. We find that the effect of nonlocal scale $M$ in the theory is to suppress the vacuum tunneling rate from false to true vacuum compared to the standard local bouncing scenario. Likewise, we show that as we move further away from the bubble wall, the effects of nonlocality gets reduced and this suppression is significant only around the wall of the nucleated bubble. From our investigations, we conclude that the main effect is due to the fact that the nonlocality smears the solution of the local bubble profile. However, the energy of the bubble wall remains unaffected by the microscopic nonlocal behavior of the theory in the thin-wall approximation. We also discuss the cases for Lee-Wick theories and applications of our result to cosmology.}


\arxivnumber{2203.04438}

\begin{document}

\maketitle
\flushbottom

\section{Introduction}
\label{intro}

An inspiration from string theory often entails that strings are nonlocal objects \cite{Witten:1985cc,Freund:1987kt,Freund:1987ck,Brekke:1987ptq,Frampton:1988kr,Kostelecky:1988ta,Kostelecky:1989nt,Tseytlin:1995uq,Siegel:2003vt,Biswas:2004qu,Calcagni:2013eua,Calcagni:2014vxa}. However, the original motivation for weakly nonlocal field theory was to formulate a novel pathway for possible UV-regularized theories as inspired from string field theory \cite{Witten:1985cc,Taylor:2003gn} in the context of weakly-coupled nonlocal Quantum Field Theory (QFT) as studied in Refs. \cite{Moffat:1990jj, Evens:1990wf, Kleppe:1991rv, Tomboulis:1997gg, Clayton:2001xz, Biswas:2011ar, Modesto:2011kw, Modesto:2012ga, Modesto:2012ys, Modesto:2013oma, Biswas:2013cha, Modesto:2014xta, Modesto:2014lga, Talaganis:2014ida, Modesto:2015lna, Modesto:2015foa, Tomboulis:2015gfa, Modesto:2017hzl, Hashi:2018kag, Buoninfante:2018mre, Modesto:2021ief, Modesto:2021okr}. This became an alternative way to address the divergence and the hierarchy problems\footnote{In local QFT, renormalization leads to ultraviolet (UV) divergences, which in the Standard Model (SM) of particle physics, a similar issue infamously known as the hierarchy problem where the Electroweak (EW) scale or the Higgs mass and Planck scale (M$_{pl}$) are separated by 18 \textit{orders-of-magnitude} apart leading to huge fine-tuning in the SM \cite{Haber:1984rc} which for example needs several new particles at the or near the EW scale, like in supersymmetry (SUSY) etc. but no new physics results from the Large Hadron Collider (LHC) has been obtained, so alternative solutions are actively being pursued.} in the Standard Model (SM) by generalizing the kinetic energy operators (which is second in order) to an infinite series of higher-order derivatives that are suppressed due the scale of nonlocality ($M$) \cite{Krasnikov:1987yj,Moffat:1990jj,Biswas:2014yia}. Such modifications to the kinetic energy sector via introducing higher-derivatives are free of ghosts\footnote{The unitarity issues are well addressed and understood in Euclidean space (imaginary time/energy) and using Cutkosky rules, the results obtained there-in can be analytically continued to Minkowski spacetime (real time/energy) via the Pius-Sen prescription \cite{Pius:2016jsl,Briscese:2018oyx,Chin:2018puw,Briscese:2021mob,Koshelev:2021orf}.} \cite{Buoninfante:2018mre}, and also cures the vacuum instability problem of the Standard Model (SM) Higgs \cite{Ghoshal:2017egr} as have been studied by one of the authors. It was found out that the $\beta$-functions, reaches a conformal limit\footnote{For other conformal theory applications to cosmology, see \cite{Barman:2021lot, Barman:2022njh} for dark matter and \cite{Ghoshal:2022hyc, Ghoshal:2022qxk} for inflation, \cite{Dasgupta:2022isg, Ghoshal:2020vud} for baryogenesis and detectable gravitational waves, just for some examples.} resolving the issue of Landau-poles \cite{Ghoshal:2020lfd}. Thus, via capturing the infinite derivatives by exponential of an entire function we get softened ultra-violate (UV) behavior in the most desirable manner, without introducing any new degrees of freedom that contribute to the particle mass spectrum, as they contain no new poles in the propagators (see Refs. \cite{Buoninfante:2018gce,Ghoshal:2018gpq} for astrophysical implications, dimensional transmutation and dark matter phenomenology in these theories.). An estimated bound on the scale of nonlocality from such observations is $M \geq O(10)$ TeV (Refs. \cite{Biswas:2014yia,Ghoshal:2018gpq}). The nonlocal theory further provided interesting implications in proton decay and Grand Unified Theories (GUT) recently in Ref. \cite{Krasnikov:2020kgh}, as well as in braneworld models, \cite{Nortier:2021six}. On the other hand, the nonperturbative strongly-coupled regimes and exact $\beta-$functions, and conditions of confinement, in higher-derivative nonlocal theories are being actively investigated by one of the authors \cite{Frasca:2020ojd,Frasca:2020jbe,Frasca:2021iip,Frasca:2022duz,Frasca:2022lwp}. The results obtained so far shows that the effect of the nonlocality in the strong coupling limits is to dilute any mass gap in the UV that the system generates. In context to gravity theories, i.e. string-inspired nonlocal extensions of the Einstein gravity, it has been shown that one can get rid of classical singularities, such as black hole singularities~\cite{Biswas:2011ar,Biswas:2013cha,Frolov:2015bia,Frolov:2015usa,Koshelev:2018hpt,Koshelev:2017bxd,Buoninfante:2018xiw,Cornell:2017irh,Buoninfante:2018rlq,Buoninfante:2018stt,Abel:2019zou,Buoninfante:2020ctr}, cosmological singularities~\cite{Biswas:2005qr,Biswas:2006bs,Biswas:2010zk,Biswas:2012bp,Koshelev:2012qn,Koshelev:2018rau,Kolar:2021qox}, and topological defects \cite{Calcagni:2008nm,Boos:2018bxf,Boos:2018bhd,Kolar:2020bpo,Boos:2020kgj,Kolar:2020max,Kolar:2021uiu}.\\


Understanding tunneling and investigating the false vacuum of a scalar field theory and the process of its decays via
the nucleation of true vacuum bubbles have been an important direction in field theory with huge implications in the analysis of phase transitions and Gravitational Waves (GW) signals in cosmology, however first studied in detail decades ago~\cite{Kobzarev:1974cp,Coleman:1977py}. As we understand from the available studies, the bubble nucleation rate is exponentially suppressed,
with the exponent being twice the Brillouin-Kramers-Wentzel (WKB) barrier penetration integral or, equivalently, one can say the Euclidean action of the bounce solution to the Euclidean field equations~\cite{Coleman:1977py}.  Via the implementation of path integral methods to estimate the energy density of the false vacua, the pre-exponential factor of the decay rate can be estimated~\cite{Callan:1977pt}, which is just the $[\det S_E''(\phi_{\rm bounce})]^{-1/2}$, where $S_E''$ denotes the second functional
variation of the Euclidean action under consideration\footnote{Coleman's method does not hold to be true in non-perturbative regimes which require a different strategy; such studies have been of recent interest, see Refs. \cite{Croon:2021vtc,Frasca:2022kfy}.}. However, an essential ingredient of this calculation is the fact that $S_E''(\phi_{\rm bounce})$ has a single negative eigenvalue, which arises from a mode corresponding to expansion or contraction of the bounce. Because of the square root, this negative mode leads to a factor of $i$ that makes the false vacuum energy complex, with its imaginary part interpreted in terms of a rate of decay by bubble nucleation; the bounce with the lowest action has one, and only one, negative mode as shown in Ref.~\cite{Coleman:1987rm}. Recently false vacuum decay estimates have been investigated in the nonperturbative
regimes by one of the authors and others and has been a topic for great interest
with respect to GW aspects \cite{Croon:2021vtc, Frasca:2022kfy, Calcagni:2022tls}.\\

Coleman and De Luccia (CDL)~\cite{Coleman:1980aw} proposed that this formalism can be extended to include gravitational effects by adding a Euclidean Einstein-Hilbert term to the action.  Although they did not address the issue of the pre-factor, it seems natural to assume that it should be analogous to that for the non-gravitational case, and that the dominant bounce should again have a single negative mode, however the analysis complicated as in gravity one also needs to take into account the
fluctuations about the bounce which enjoy a gauge freedom, corresponding
to the invariance under coordinate transformations of the underlying
theory, leading to some constraints that must be imposed on the possible fluctuations.\\

Motivated by the nonlocal UV-completion of the QFT and the standard model, the goal of this study is to compute the bubble nucleation and tunneling rate in a toy model with a scalar field with 2 slightly non-degenerate vacua, such that one may perform the thin-wall approximation. In order to get an approximate solution for the Coleman bounce with weak nonlocality\footnote{Other works treating the effects of time quantum mechanical aspects in nonlocal theories, cf. \cite{Buoninfante:2017kgj, Buoninfante:2017rbw, Buoninfante:2019teo} for instance.}, we use the ``diffusion'' method of Refs.~\cite{Calcagni:2007ef, Calcagni:2008nm, Calcagni:2009jb, Cembranos:2016dll, Calcagni:2018lyd, Calcagni:2018fid}, and compare with the local case.

\newpage

The paper is organized as follows. In Section~\ref{review}, we remind the main features of a string-inspired nonlocal theory for a scalar field. In Section~\ref{model}, we present the toy model under study, and we express the vacuum decay rate and the energy of the bubble wall in the thin-wall approximation. In Section~\ref{sec_soliton_sol}, we study the nonlocal soliton solution, which is used to compute the Coleman bounce in the thin-wall approximation. In Section~\ref{conclusion}, we summarize our results.

\medskip

\section{Short review: String-inspired nonlocal scalar field theory}
\label{review}

Typically, the action for the string-inspired nonlocal theory given by \cite{Calcagni:2009jb, Buoninfante:2018mre}:
\begin{equation} \label{Action}
S = \int d^4 x\ \left(\frac{1}{2} \phi e^{f(\Box)}(\Box-m^2)\phi -\frac{\lambda}{4!}\phi^{4}\right)
\end{equation} 
where the normalization of $\phi$ is so chosen such that the residue at the $p^2=-m^2$ pole is unity in this case, and $\Box= \eta_{\mu\nu}\partial^{\mu}\partial^{\nu}$ $(\mu, \nu=0,1,2,3$) with the convention of the metric signature $(-,+,+,+)$, $m$ is the mass of the scalar particle, and $M$ is the energy scale of the nonlocality which is usually taken to be below the Planck scale. The most important thing to note here is the fact that the kinetic energy terms are generalized with higher-derivatives suppressed by the nonlocal energy scale $M$, whilst the scalar self-interaction is the typical quartic potential one. This action is reduced into the standard local field theory taking the limit of $M \to \infty$\footnote{Regarding the nonlocal form-factor $e^{f(\Box)}$, several possibilities have been popularly considered in the existing literature \cite{Edholm:2016hbt}. Here, however, for simplicity, we will use  $e^{-\frac{\Box}{M^2}}$. Nonetheless, the essential criterion is that the nonlocal factor is an entire function without zeros, which allows us to avoid the appearance of possible ghosts in the physical spectrum of the theory.}.\\


In Euclidean space ($p^0\rightarrow ip_{E} ^0$), the propagator is given by,
\be \label{Prop}
\Pi(p^2)=-\frac{ie^{f(-p_E^2)}}{p_E^2+m^2}
\ee
and the vertex factor is, as usual, given by $-i\lambda$.
It can be clearly noted that the nonlocal extensions of the local theory leads to the exponential suppression of the propagators for $p_E^2 > M^2$ region. What this essentially means is that the quantum corrections in the ultraviolet (UV) will be frozen at energies higher than $M$ (basically all the $\beta-$functions vanish beyond the scale of nonlocality $M$ thereby making the theory reach an asymptotically conformal limit in the UV) \cite{Ghoshal:2017egr,Ghoshal:2020lfd} which is a desirable feature in any QFT. In the other popular form factor considered in Refs.~\cite{Modesto:2015lna,Modesto:2015foa}, one can even weaken the interaction above $M$ and achieve asymptotic freedom (the coupling goes to zero in the UV).\\

An equivalent description of the action in Eq.~\eqref{Action} is the usual local Klein-Gordon kinetic operator via the following field re-definition \cite{Buoninfante:2018mre}:
\begin{equation}
\begin{array}{rl}
\tilde{\phi}(x)= & \displaystyle e^{-\frac{1}{2}f(\Box)}\phi(x)\\
= & \displaystyle \int d^4y \mathcal{F}(x-y)\phi(y),
\end{array}
\label{42}
\end{equation}
where $\mathcal{F}(x-y):=e^{-\frac{1}{2}f(\Box)}\delta^{(4)}(x-y)$ is the kernel of the differential operator $e^{-\frac{1}{2}(\Box)}$. One gets the following action with an added source term $j(x)$:
\begin{equation}
S=\frac{1}{2}\int d^4x \left(\tilde{\phi}(x)(\Box-m^2)\tilde{\phi}(x) -\int d^4x \frac{ \lambda }{4!}\left(e^{\frac{1}{2}f(\Box)}\tilde{\phi}(x)\right)^4
+j(x)e^{\frac{1}{2}f(\Box)}\tilde{\phi}(x)\right) .
\label{43}
\end{equation}

\section{Model of the bounce}
\label{model}

In this section we show the model of the bounce via the scalar field in local theory and then 
we show how the picture changes when nonlocality is introduced.

\subsection{Weakly nonlocal scalar field theory}
We consider a scalar field theory defined on a 4D Euclidean space $\mathbb{R}^4$:
\begin{equation}
ds^2 = \delta_{\mu \nu} \, dx^\mu dx^\nu \, , \ \ \ \delta_{\mu \nu} = (+1, +1, +1, +1) \, .
\end{equation}
Indeed, since the usual Wick rotation is not defined in string-inspired nonlocal theories, one must define the theory on an Euclidean space and then analytically continue to Minkowski spacetime only for the asymptotic states because of unitary issues \cite{Pius:2016jsl,Briscese:2018oyx,Chin:2018puw,Briscese:2021mob,Koshelev:2021orf}. In other words, the Lorentzian causal structure of spacetime emerges only at scales larger than the nonlocal length scale $\ell=1/M$ \cite{Carone:2016eyp, Buoninfante:2018mre}.\\

We introduce a real scalar field $\phi(x)$, whose potential is even under the $\mathbb{Z}_2$ transformation ($\phi \mapsto -\phi$):
\begin{equation}
V_0 (\phi) = V_\mu(\phi) + V_I(\phi) + \Lambda_C \, ,
\end{equation}
with the mass term ($\mu > 0$):
\begin{equation}
V_\mu(\phi) = - \dfrac{\mu^2}{2} \phi^2 \, ,
\end{equation}
the quartic (interaction) term ($\lambda > 0$):
\begin{equation}
V_I(\phi) = \dfrac{\lambda}{4!} \, \phi^4 \, ,
\end{equation}
and the "cosmological constant" term $\Lambda_C$. The 2 symmetric minima of $V_0 (\phi)$ are
\begin{equation}
\phi_\pm = \pm v = \pm \sqrt{\dfrac{6}{\lambda}} \mu \, ,
\end{equation}
and the "cosmological constant" $\Lambda_C$ is defined such that
\begin{equation}
V_0 (\pm v) = 0 \ \ \ \Rightarrow \ \ \ \Lambda_C = \left( \dfrac{\mu v}{2}\right)^2 > 0 \, .
\end{equation}
Let us focus on the Euclidean kinetic energy operator, which is now modified via infinite series of higher-order derivatives:
\begin{equation}
(\square + \mu^2) e^{- \ell^2 (\square + \mu^2)} \, ,
\end{equation}
with the Euclidean d'Alembertian
\begin{equation}
\square = \delta^{\mu\nu} \, \partial_\mu \partial_\nu \, ,
\end{equation}
such that the parameter $\ell$ interpolates between a local and a nonlocal field theory:
\begin{itemize}
\item in a local field theory: $\ell = 0$.
\item in a string field theory-inspired nonlocal field theory: $\ell \neq 0$.
\end{itemize}
Let us define the parameter $\omega = (\mu \ell)^2$, such that $\sqrt{\omega}$ measures the separation between the energy scales $\mu$ and $M$.\\

We wish to investigate the bubble nucleation when one of the two vacua is metastable through quantum effects, and decay to the stable one by barrier penetration. Therefore, one needs to introduce an additional term $V_\epsilon(\phi)$ to the potential $V_0 (\phi)$, which is asymmetric with respect to the $\mathbb{Z}_2$ transformation, and thus leaves us with a certain degeneracy between the two vacua ($\phi_+ \neq \phi_-$). The magnitude of $V_\epsilon(\phi)$ is controlled by a small parameter
\begin{equation}
\epsilon = V(\phi_+) - V(\phi_-) > 0 \, ,
\end{equation}
such that
\begin{equation}
V_\epsilon(\phi) = V(\phi) - V_0(\phi) = \mathcal{O}(\epsilon) \, ,
\end{equation}
where $V(\phi)$ is the total scalar potential. Now, the Euclidean action of this scalar field theory is thus
\begin{equation}
S = \int d^4 x \ \mathcal{L} \, .
\end{equation}
with the Euclidean Lagrangian
\begin{equation}
\mathcal{L} = - \dfrac{1}{2} \, \phi \, (\square + \mu^2) e^{- \ell^2 (\square + \mu^2)} \phi + U(\phi) \, ,
\end{equation}
where the total potential is now
\begin{equation}
U(\phi) = U_0(\phi) + V_\epsilon(\phi) , \ \ \ U_0(\phi) = V_0(\phi) - V_\mu(\phi) \, .
\end{equation}
One can rescale the field $\phi \mapsto e^{\omega/2} \phi$, such that the quartic coupling is also rescaled: $\lambda \mapsto e^{-2 \omega} \lambda$. It is also useful to define the rescaled VEV $v \mapsto e^{\omega} v$. 
By applying Hamilton's principle, one arrives at the Euler-Lagrange equation for $\phi$:
\begin{equation}
(\square + \mu^2) e^{- \ell^2 \square} \phi = \partial_\phi U (\phi) \, .
\label{EL_eq_scalar}
\end{equation}

First, let us assume that a bounce solution \cite{Coleman:1977py} to Eq.~\eqref{EL_eq_scalar} exists also in the nonlocal case, at least for some region of the parameter space (we will come back to this assumption at the end of Section~\ref{sec_soliton_sol}). At the semi-classical level, the probability of decay of the false vacuum per unit time per unit volume takes the form \cite{Coleman:1977py}
\begin{equation}
\Gamma/V = A e^{-B/\hslash} \left[ 1 + \mathcal{O}(\hslash) \right] \, ,
\end{equation}
where $\hslash$ is the reduced Planck constant\footnote{In the rest of the article, we work in the natural unit system, where $\hbar = 1$ and the speed of light in vacuum is $c=1$.}. In the same spirit as in Ref.~\cite{Coleman:1977py}, one of the goals of this article is to compute the coefficient $B$ for $\epsilon \ll 1$. This coefficient $B$ is given by the on-shell Euclidean action
\begin{equation}
B = S(\phi) - S(\phi_+) \, ,
\end{equation}
where $\phi(x)$ is a bounce solution of Eq.~\eqref{EL_eq_scalar}, i.e. satisfying the following properties:
\begin{itemize}
\item The bounce $\phi(x)$ is not constant;
\item The bounce $\phi(x)$ approaches the false vacuum $\phi_+$ at Euclidean infinity;
\item The Euclidean action of the bounce $\phi(x)$ is less or equal to that of any other solution satisfying the two previous properties.
\end{itemize}
Under the assumption that the bounce $\phi(x)$ has a $O(4)$ symmetry, one introduces hyperspherical coordinates, such that $\phi(x) = \phi(\rho)$, where $\rho \geq 0$ is the radial direction coordinate:
\begin{equation}
\rho = \sqrt{\delta^{\mu\nu} \, x_\mu x_\nu} \, .
\end{equation}
The Euclidean d'Alembertian applied on $\phi(\rho)$ gives us
\begin{equation}
\square \phi (\rho) = \left( \partial_\rho^2 + \dfrac{3}{\rho} \, \partial_\rho \right) \phi (\rho) \, .
\label{box_rho}
\end{equation}
Therefore, the coefficient $B$ is the on-shell Euclidean action:
\begin{align}
B = S &= 2 \pi^2 \int_{-\infty}^{+\infty} \rho^3 d\rho \ \mathcal{L} , \nonumber \\
&= 2 \pi^2 \int_{-\infty}^{+\infty} \rho^3 d\rho \left[ - \dfrac{1}{2} \, \phi \, \partial_\phi U(\phi) + U(\phi) \right] \, .
\label{exact_B}
\end{align}
Thus, we arrive a generic expression for $B$, however in order to proceed we need to make some assumptions and compute the quantities in those limits.

\subsection{Thin-wall approximation}

As shown in Ref.~\cite{Coleman:1977py} and what is famously known as the thin-wall approximation, which is usually valid when $\epsilon \ll 1$, one considers the bubble nucleates at a large time $\rho = R \gg 1/\mu, \ \ell$, $\rho$ traverses the real line, and the bounce $\phi(\rho)$ goes monotonically from $\phi_-$ to $\phi_+$. Here, the bounce looks like a large 4D bubble of radius $R$, with a thin wall separating the two vacua $\phi_\pm$. At the end of the computations, it is important to check if one gets $R \gg 1/\mu, \ \ell$ when $\epsilon \ll 1$. In the thin-wall approximation, $\phi(\rho)$ has the form:
\begin{equation}
\phi(\rho) =
\begin{cases} 
- \tilde{v} & \text{if} \ \rho \ll R \, , \\
\phi_w(\rho - R) & \text{if} \ \rho \sim R \, , \\
+ \tilde{v} & \text{if} \ \rho \gg R \, .
\end{cases}
\label{bounce_sol}
\end{equation}
Indeed, far away from the bubble wall, the field theory is in the vacuum $\phi_\pm$, and the infinite series of derivatives play no role, since $\phi(\rho)$ varies only in the thin-shell of the wall, where $\rho \sim R$, where $\phi(\rho)$ can be approximated by a function $\phi_w(\rho-R)$. For $\rho \sim R$, one can solve the Euclidean Euler-Lagrange equation \eqref{EL_eq_scalar}, with the approximate symmetric potential $U(\phi) \sim U_0(\phi)$, and the approximate kinetic operator
\begin{equation}
(\square + \mu^2) e^{- \ell^2 \square} \phi_w \sim (\partial_\rho^2 + \mu^2) e^{- \ell^2 \partial_\rho^2} \phi_w \, ,
\label{approx_box}
\end{equation}
with the hypothesis that the viscous damping term in the d'Alembertian \eqref{box_rho} can be neglected ($\square \phi_w \sim \partial_\rho^2 \phi_w$), which will be checked at the end of the computation. If one performs the change of variable $r = \mu (\rho - R)$), and one defines $\varphi(\omega, r) = \phi_w(r) / \tilde{v}$, the Euler-Lagrange equation for $\phi_w(r)$ is approximated by the equation of a soliton for $\varphi(\omega, r)$:
\begin{equation}
(\partial_r^2 + 1) e^{- \omega \partial_r^2} \varphi(\omega, r) - \varphi^3(\omega, r) = 0 \, ,
\label{soliton_eq}
\end{equation}
whose solution is odd with respect to $r=0$, thus for $R \gg 1/\mu, \ \ell$, it is clear that the viscous damping term neglected in Eq.~\eqref{approx_box} will give a small perturbation to the approximate soliton profile. Eq.~\eqref{soliton_eq} can be obtained from Hamilton's principle applied to the Euclidean action
\begin{equation}
\int_{-\mu R}^{+ \infty} \dfrac{dr}{\mu} \ \mathcal{L}_w \underset{\mu R \gg 1}{\simeq} \int_{-\infty}^{+ \infty} dr \ \mathcal{L}_w \, ,
\end{equation}
with
\begin{equation}
\mathcal{L}_w = - \dfrac{1}{2} \, \varphi \, (\partial_r^2 + 1) e^{- \omega \partial_r^2} \varphi + \varphi^4 + \Lambda_C \ \, .
\end{equation}
The on-shell Euclidean action for the bounce near the bubble wall is then
\begin{equation}
S_w(\omega) \simeq \dfrac{\Lambda_C}{\mu} \int_{-\infty}^{+\infty} dr \left[ 1 - \varphi^4(\omega, r) \right] \, .
\end{equation}

In the thin-wall approximation, from Eqs.~\eqref{bounce_sol} and \eqref{exact_B}, the coefficient $B$ is given by
\begin{equation}
B = B_i + B_w + B_o \, ,
\end{equation}
where the contribution inside the bubble is
\begin{equation}
B_i = - \dfrac{\pi^2}{2} R^4 \epsilon \, ,
\end{equation}
the contribution of the wall is
\begin{equation}
B_w = 2 \pi^2 R^3 S_w \, ,
\end{equation}
the contribution outside the bubble is
\begin{equation}
B_o = 0 \, ,
\end{equation}
and the value of $R$ is given by the one which minimizes the on-shell action $B$:
\begin{equation}
\dfrac{dB}{dR} = 0 \ \ \ \Rightarrow \ \ \ R = \dfrac{3 S_w}{\epsilon} \, .
\label{R_value}
\end{equation}
One can thus see that as long as $S_w(a) > 0$, we can choose $\epsilon$ sufficiently small in order to justify the thin-wall approximation, such that $R \gg 1/\mu, \ \ell$, which justified a posteriori all the previous approximations. In this case, one obtains
\begin{equation}
B = \dfrac{27 \pi^2 S_w^4}{2 \epsilon^3} \, ,
\label{B_value}
\end{equation}
which has the same qualitative form in both the local and nonlocal cases. In what follows, one thus needs to determine the quantity $S_w(\omega)$. In order to compare the on-shell action near the bubble wall $S_w$ for the nonlocal versus local field theory, it is useful to introduce a function
\begin{equation}
Q(\omega) = \dfrac{S_w(\omega)}{S_w(0)} \, ,
\label{Q}
\end{equation}
such that with Eq.~\eqref{B_value} becomes:
\begin{equation}
B(\omega) = B(0) \, Q^4(\omega) \, ,
\end{equation}
and the thin-wall approximation means that we require $Q(\omega) > 0$. Note that we choose to normalize with a local model which have the same quartic coupling than the nonlocal one, since the quartic coupling has been rescaled ($\lambda \mapsto e^{-2 \omega} \lambda$).\\

Before computing $Q(\omega)$ in the section~\ref{sec_soliton_sol}, it is interesting to estimate the energy of the wall $E_w$. Outside the bubble wall, where nonlocality plays no role, one can perform an analytical continuation of the bounce in Minkowski space-time ($x^0 = it$):
\begin{equation}
\phi \left( t, \vec{x} \right) = \phi \left( \rho = \sqrt{\left| \vec{x} \right|^2 - t^2} \right) \, .
\end{equation}
In the limit where the bubble wall can be considered as a thin-shell, a section of this bubble wall at rest carries energy $S_w$, per unit area. Therefore, a section of the wall expands with a velocity 
\begin{equation}
v = \dfrac{d \vec{x}}{dt} = \dfrac{\sqrt{\left| \vec{x} \right|^2 - R^2}}{\vec{x}} \, ,
\label{v_value}
\end{equation}
and carries an energy
\begin{equation}
\dfrac{S_w}{\sqrt{1-v^2}}
\end{equation}
per unit area. At a time when the radius of the bubble is $|\vec{x}|$, the energy of the wall is then
\begin{align}
E_w &= \dfrac{4 \pi \left| \vec{x} \right|^2 \, S_w}{\sqrt{1-v^2}} \nonumber \\
&= \dfrac{4 \pi \left| \vec{x} \right|^3 \, S_w}{R} \nonumber \\
&= \dfrac{4 \pi \epsilon \left| \vec{x} \right|^3}{3} \, ,
\label{wall_energy}
\end{align}
where one used Eq.~\eqref{v_value} to get the second and third lines, respectively. At the end, one sees that $E_w$ depends only on the parameter $\epsilon$, which controls the magnitude of the asymmetry of the scalar potential, and not on the microscopic scales $\mu$ and $M$, in the thin-wall approximation. It is not so surprising because the scales $\mu$ and $\Lambda$ mainly determines the width of the soliton-like profile of the bubble wall (cf. Eq.~\eqref{soliton_eq}), which we have considered as negligible in the computation.

\section{Solution of the soliton}
\label{sec_soliton_sol}

\subsection{Local solution}
\begin{figure}[h]
\begin{center}
\includegraphics[width=13cm]{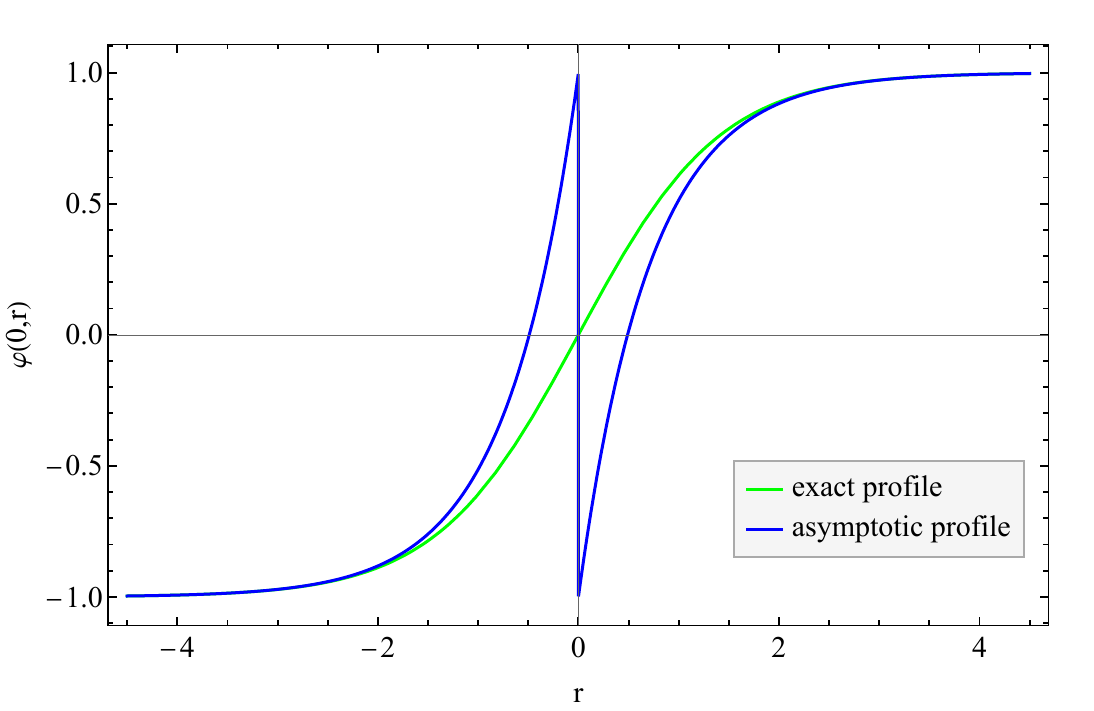}
\end{center}
\caption{ \it In green: exact profile $\varphi(0, r)$ (Eq.~\eqref{exact_local}) of the local soliton. In blue: asymptotic profile $\varphi(0, r)$ (Eq.~\eqref{asympt_local}) of the local soliton ($|r| \gg 1$).}
\label{plot_local_soliton}
\end{figure}

In this subsection, we will determine $\varphi(0, r)$ which is the toy model of Ref.~\cite{Coleman:1977py}. The exact solution of the local soliton equation is (cf. Fig.~\ref{plot_local_soliton})
\begin{align}
\varphi(0, r) = \tanh \left( \dfrac{r}{\sqrt{2}} \right) \, .
\label{exact_local}
\end{align}
Since the soliton equation gives the profile of the bubble wall in the approximation $|r| \gg 1$, one can only keep an asymptotic expansion (cf. Fig.~\ref{plot_local_soliton}):
\begin{equation}
\varphi(0, r) \underset{|r| \ll 1}{\sim} \text{sgn}(r) \left( 1-2 e^{- \sqrt{2} |r|} \right) \, ,
\label{asympt_local}
\end{equation}
with the sign function defined as
\begin{equation}
\sgn(r) =
\begin{cases} 
- 1 & \text{if} \ r < 0 \, , \\
0 & \text{if} \ r = 0 \, , \\
+ 1 & \text{if} \ r > 0 \, .
\end{cases}
\end{equation}
The on-shell action $S_w(0)$ for the local problem is
\begin{equation}
S_w(0) = \dfrac{2 \sqrt{2}}{3} \mu v^2 \, ,
\end{equation}
for both the exact and asymptotic profiles of the local soliton.

\subsection{Nonlocal solution}
In order to solve the nonlocal soliton equation, one can use the ``diffusion'' method of Refs.~\cite{Calcagni:2007ef, Calcagni:2008nm, Calcagni:2009jb, Cembranos:2016dll, Calcagni:2018lyd, Calcagni:2018fid} to get an approximate solution for $|r| \gg 1$. One considers that the field $\varphi(\omega, r)$, where $\omega$ is an auxiliary coordinate, satisfies the diffusion equation (with $\beta > 0$):
\begin{equation}
\left( \partial_r^2 - \dfrac{\partial_\omega}{\beta} \right) \varphi (\omega, r) = 0
 \ \ \ \Rightarrow \ \ \ 
\varphi(\omega, r) =  e^{\beta \omega \partial_r^2} \varphi(0, r) \, .
\label{diffusion_eq}
\end{equation}
 
\newpage

\begin{figure}[h]
\begin{center}
\includegraphics[width=13cm]{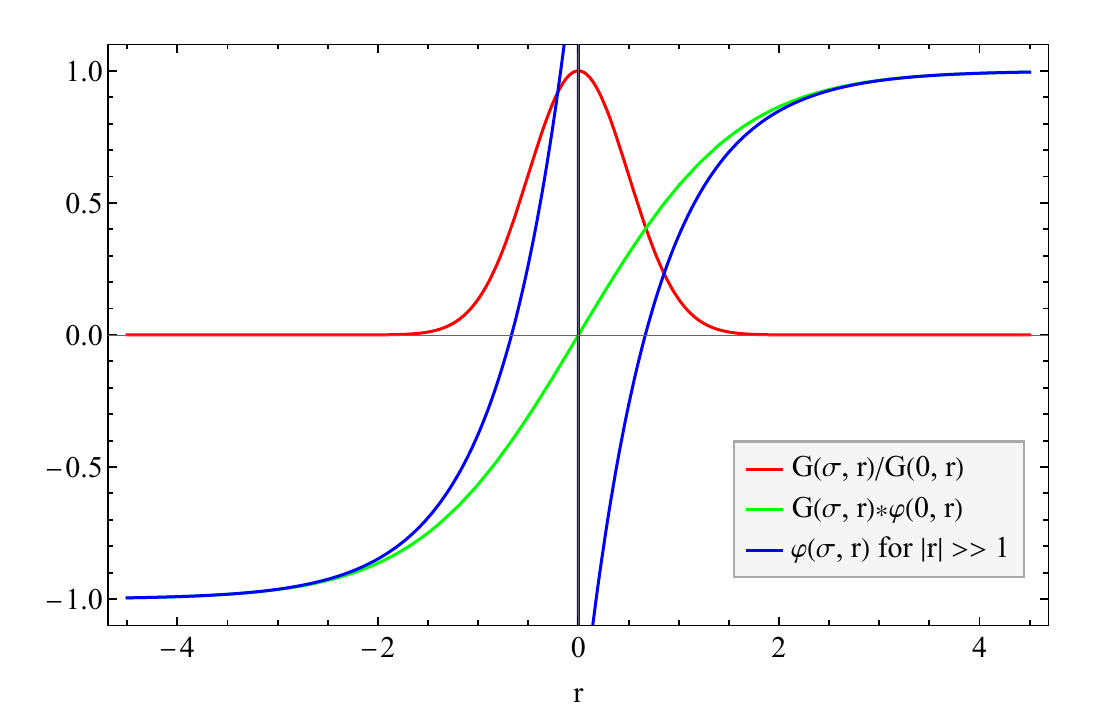}
\end{center}
\caption{ \it For $\sigma = 1/2$, in red: the gaussian function \eqref{Gaussian}; in green: the exact convolution product \eqref{exact_conv} with the exact local soliton \eqref{exact_local}; in blue: the asymptotic nonlocal profile \eqref{asympt_nloc_soliton} of the soliton $\varphi(\sigma, r)$ with $|r| \gg 1$.}
\label{plot_nloc_soliton_0}
\end{figure}

\begin{figure}[h]
\begin{center}
\includegraphics[width=13cm]{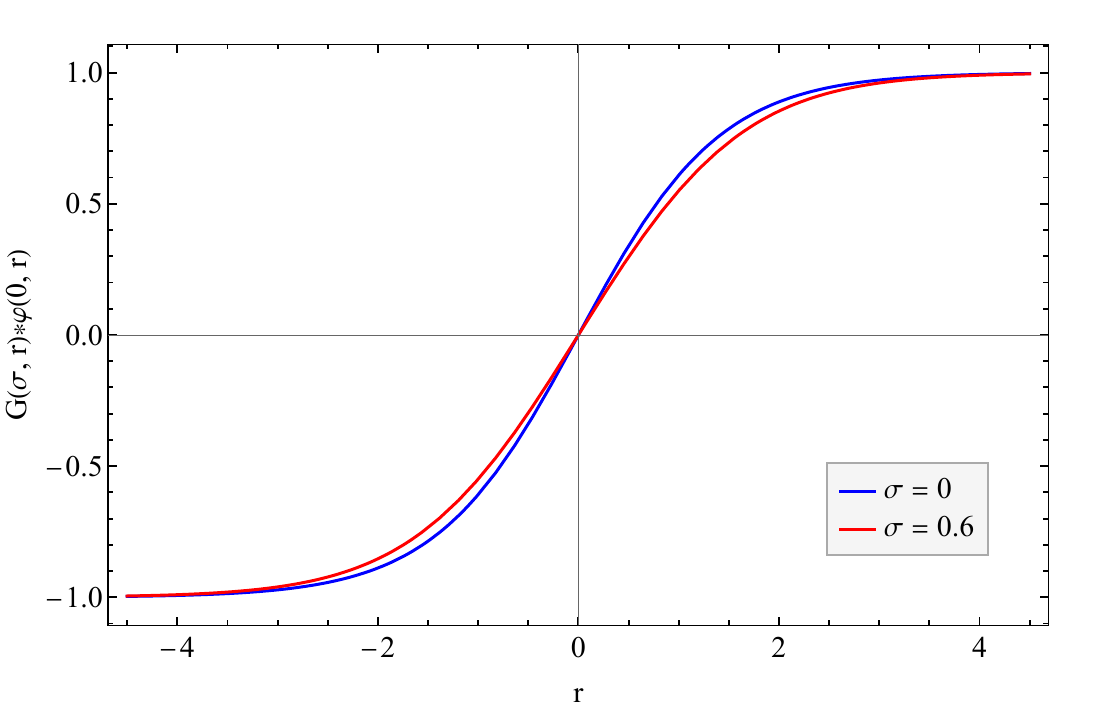}
\end{center}
\caption{ \it In (blue, red), exact convolution product \eqref{exact_conv} with the exact local soliton \eqref{exact_local} for $\sigma = (0, 0.6)$.}
\label{plot_nloc_soliton_0_sigma}
\end{figure}

\noindent By definition:
\begin{itemize}
\item $\varphi(0, r)$ is the profile of the soliton in the local case ($\omega = 0$);
\item $\varphi(\omega, r)$ is the profile of the soliton in the nonlocal ($\omega \neq 0$).
\end{itemize}
One introduces the normalized Gaussian function
\begin{equation}
G(\sigma, r) = \dfrac{1}{\sqrt{2 \pi} \sigma} \, \exp \left( - \dfrac{r^2}{2 \sigma^2} \right) \, ,
\end{equation}
such that it defines the heat kernel as
\begin{align}
G(\sigma, r - r') &= \exp \left( \dfrac{\sigma^2}{2} \, \partial_r^2 \right) \delta(r - r') \, .
\label{Gaussian}
\end{align}
One can then rewrite
\begin{align}
\varphi(\omega, r) &= e^{\beta \omega \partial_r^2} \varphi(0, r) \, , \nonumber \\
&= \int dr' \ e^{\beta \omega \partial_r^2} \delta(r - r') \, \varphi(0, r') \, .
\end{align}
By performing the change of variable $\sigma^2 = 2 \beta \omega$, one gets
\begin{align}
\varphi(\sigma, r) &= \int dr' \ G(\sigma, r-r') \, \varphi(0, r') \, , \nonumber  \\
&= G(\sigma, r) * \varphi(0, r)  \, ,
\end{align}
where $*$ is the convolution product\footnote{See Refs.~\cite{Heredia:2021pxp,Kolar:2022kgx} for weakly nonlocal theories in convolutional form.} with respect to $r$. The new variable $\sigma \propto \sqrt{\omega}$ and is thus related to the separation of the scales $\mu$ and $M$. One can compute numerically the convolution of the exact soliton profile \eqref{exact_local} with the Gaussian function \eqref{Gaussian}, where one can see in Figs.~\ref{plot_nloc_soliton_0} and \ref{plot_nloc_soliton_0_sigma} that nonlocality ($\sigma \neq 0$) smears the behavior of the local soliton near the kink when $\sigma$ increases: the soliton width is increased by the smearing effect of nonlocality. This effect is already known for zero-thickness topological defects \cite{Calcagni:2008nm,Boos:2018bxf,Boos:2018bhd,Kolar:2020bpo,Boos:2020kgj,Kolar:2020max,Nortier:2021six}. Far from the wall, the nonlocal soliton smoothly recovers the asymptotic behavior of the local one. Again, the effect of the infinite series of derivatives is important when spacetime variations of the field are important, such as near the bubble wall.

\newpage

\newpage

\begin{figure}[h]
\begin{center}
\includegraphics[width=13cm]{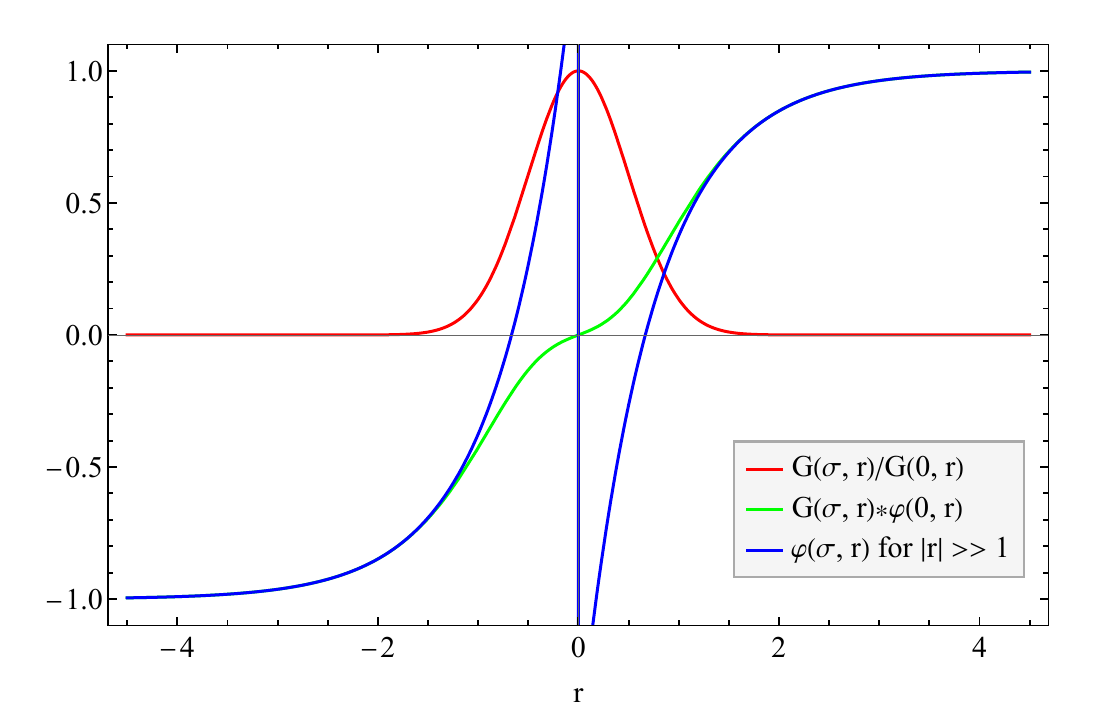}
\end{center}
\caption{ \it For $\sigma = 1/2$, in red: the Gaussian function \eqref{Gaussian}; in green: the exact convolution product \eqref{exact_conv} with the asymptotic local soliton \eqref{asympt_local}; in blue: the asymptotic nonlocal profile \eqref{asympt_nloc_soliton} of the soliton $\varphi(\sigma, r)$ with $|r| \gg 1$.}
\label{plot_nloc_soliton}
\end{figure}

\begin{figure}[h]
\begin{center}
\includegraphics[width=13cm]{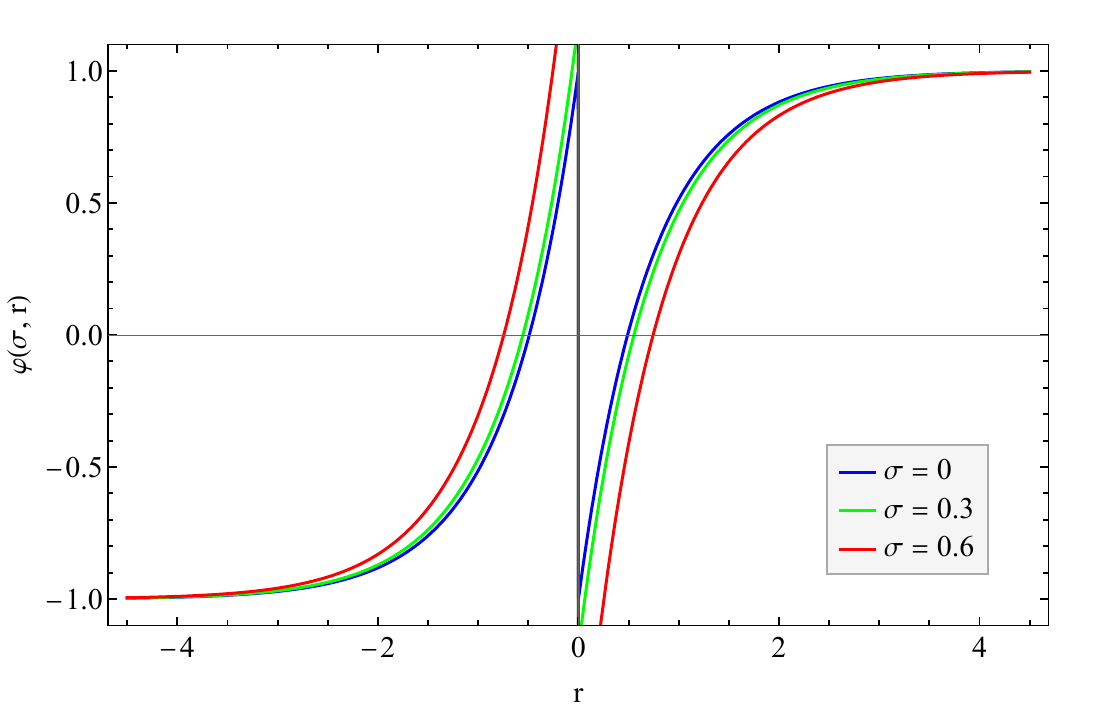}
\end{center}
\caption{ \it In (blue, green, red), asymptotic nonlocal profile \eqref{asympt_nloc_soliton} of the soliton $\varphi(\sigma, r)$ with $|r| \gg 1$ for $\sigma = (0, 0.3, 0.6)$.}
\label{plot_asympt_nloc_soliton_sigma}
\end{figure}

In this study, one needs only the asymptotic behavior for $|r| \gg 1$ to compute the coefficient $B$ in the thin-wall approximation. Therefore, from the asymptotic local solution \eqref{asympt_local} of $\varphi(0, r)$, one gets (cf. Fig.~\ref{plot_nloc_soliton}):
\begin{align}
G(\sigma, r) * \varphi(0, r) = \erf \left( \dfrac{r}{\sqrt{2} \sigma} \right) 
- e^{\sigma^2 - \sqrt{2} r} \erfc \left( \dfrac{r}{\sqrt{2} \sigma} - \sigma \right)
+ e^{\sigma^2 + \sqrt{2} r} \erfc \left( \dfrac{r}{\sqrt{2} \sigma} + \sigma \right) \, ,
\label{exact_conv}
\end{align}
with the error function
\begin{equation}
\erf(r) = \dfrac{2}{\sqrt{\pi}} \int_0^r du \ e^{-u^2} \, ,
\end{equation}
and the complementary error function
\begin{equation}
\erfc(r) = 1 - \erf(r) \, .
\end{equation}
However, only the asymptotic behavior $|r| \gg 1$ of this result is meaningful (cf. Fig.~\ref{plot_nloc_soliton}):
\begin{align}
\varphi(\sigma, r) \underset{|r| \ll 1}{\sim} \sgn(r) \left( 1 - 2 e^{ \sigma^2 - \sqrt{2} |r|} \right) \, .
\label{asympt_nloc_soliton}
\end{align}
We see on Fig.~\ref{plot_asympt_nloc_soliton_sigma} that when $\sigma \neq 0$ increases, the departure of the nonlocal soliton profile from the local one is more important near the wall, since the nonlocal scale $M$ is closer to the mass scale $\mu$ of the scalar potential, and thus the nonlocal effects are more important.

\newpage

After that, one has to determine for which value of $\beta$ the soliton profile \eqref{asympt_nloc_soliton} is the best approximation for a solution of the nonlocal soliton equation \eqref{soliton_eq}. To achieve this task, we introduce the function
\begin{align}
\forall r>0, \ E(\beta, \omega, r) &=
\begin{cases}
(\partial_r^2 + 1) \varphi(0, r) - e^{ (1-\beta) \omega \partial_r^2} \left[ e^{\beta \omega \partial_r^2} \varphi(0, r) \right]^3 & \text{if} \ \beta < 1 \, , \\
(\partial_r^2 + 1) e^{(\beta - 1) \omega \partial_r^2} \varphi(0, r) - \left[ e^{\beta \omega \partial_r^2} \varphi(0, r) \right]^3 & \text{if} \ \beta \geq 1 \, .
\end{cases}
\nonumber \\
&\underset{|r| \gg 1}{\sim}
\begin{cases}
6 \left( 1-e^{2(\beta-1) \omega} \right) e^{-\sqrt{2} r - 2(\beta-1) \omega} & \text{if} \ \beta < 1 \, , \\
6 \left( e^{2 \omega} - 1 \right) e^{- \sqrt{2} r + 2(\beta-1) \omega} & \text{if} \ \beta \geq 1 \, ,
\end{cases}
\end{align}
which evaluates the nonlocal soliton equation \eqref{soliton_eq} for the soliton profile \eqref{asympt_nloc_soliton}. It is straightforward to study the function $E(\beta, \omega, r) > 0$: there is only one minimum at $\beta = 1$ for $r \gg 1$, and one can check numerically that $E(1, \omega, r \gg 1) \ll 1$, such that $\varphi(1, r \gg 1)$ is a good approximation for the solution of Eq.~\eqref{soliton_eq}. Since the soliton is an odd function with respect to $r=0$, the same analysis holds for $r \ll -1$. In the following, one will thus take $\beta = 1$.\\

\begin{figure}[h]
\begin{center}
\includegraphics[width=13cm]{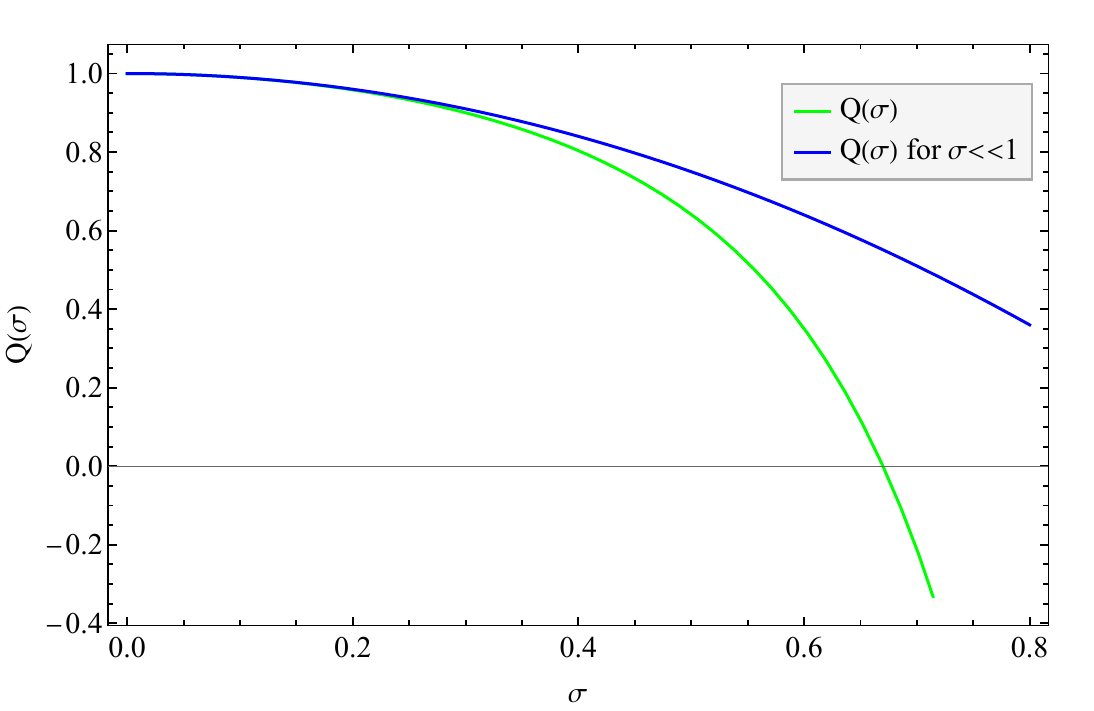}
\end{center}
\caption{ \it In green: $Q(\sigma)$ in the thin-wall approximation; in blue: $Q(\sigma)$ for $\sigma \ll 1$.}
\label{plot_Q}
\end{figure}

Now, one knows the bounce in the thin-wall approximation, which allows determining the function $Q(\sigma)$ from Eq.~\eqref{Q} (cf. Fig~\ref{plot_Q}):
\begin{align}
Q(\sigma) &= 3 - \dfrac{9 e^{\sigma^2}}{2} + 4 e^{2 \sigma^2} - \dfrac{3 e^{3 \sigma^2}}{2} \, .
\label{Q_result_nl}
\end{align}
The thin-wall approximation is valid as long as $Q(\sigma) > 0$, i.e. for $\sigma \lesssim 0.67$. One can interpret this result by the fact that for larger values of $\sigma$, the bubble wall is so smeared by nonlocality that the thin-wall approximation is meaningless, such that one needs a finer analysis to understand this region of the parameter space, which is beyond the scope of this first study. In the local case ($\sigma = 0$), one can always do a thin-wall approximation. When $M \gg \mu$,
\begin{align}
Q(\sigma) &\underset{\sigma \ll 1}{\sim} 1 - \sigma^2 \, ,
\label{local_limit}
\end{align}
and we see in Fig~\ref{plot_Q} that the behavior is closed to the local case, as expected. From Eq.~\eqref{Q_result_nl} and Fig~\ref{plot_Q}, we see that the effect of weak nonlocality is to exponentially suppress the vacuum decay rate with respect to the local theory. As discussed previously, weak nonlocality increases the bubble width, such as it is more difficult for the system to cross the potential barrier by quantum tunneling, in analogy with the well-known results in 1D quantum mechanics where the transmission amplitude of the incoming particle is exponentially suppressed when one increases the width of the potential barrier. Note that the height of the potential barrier is kept fixed, since we choose to compare the nonlocal model with a local theory which has the same effective quartic coupling below the nonlocal scale.

Finally, let us come back to our original assumption that a bounce solution to Eq.~\eqref{EL_eq_scalar} exists. In Ref.~\cite{Coleman:1977py}, Coleman shows that a bounce solution always exists in his toy scalar model. Note that if $M \gg \mu$ in our weakly nonlocal case, it is clear from the decoupling of scales that such deep-UV nonlocal scale will not spoil the existence of a bounce solution for the scalar field with a deep-IR mass scale in the potential. Now, if the nonlocal scale is closer to the mass scale of the potential with $M \geq \mu$, and that the thin-wall approximation is reliable, we showed that Eq.~\eqref{EL_eq_scalar} reduces to the soliton equation \eqref{soliton_eq}, for which we were able to give an approximate solution by the ``diffusion'' method developed in Refs.~\cite{Calcagni:2007ef, Calcagni:2008nm, Calcagni:2009jb, Cembranos:2016dll, Calcagni:2018lyd, Calcagni:2018fid}. Therefore, we showed that a bounce solution exists for $M \geq \mu$ in the region of parameter space where the thin-wall approximation is valid. The analysis of this paper does not treat the general case (for which the existence of a bounce solution may not be guaranteed), that needs to be investigated beyond the thin-wall approximation.

\section{Discussion and conclusion}
\label{conclusion}
In a simple quartic scalar field theory generalized to infinite sets of higher-order derivatives, as motivated from string field theory approaches, we investigated the false vacuum decay rate and bubble nucleation. We found that the effect of nonlocality completely changes the rate of bubble nucleation. We summarize the main findings of our paper as follows:
\begin{itemize}
    \item We investigated the Coleman bounce in string-inspired nonlocal scalar field theory in the thin-wall approximation. We used the method where nonlocality is treated as a diffusion process in an auxiliary extra dimension (cf. Eq.~\eqref{diffusion_eq}) in order to get an approximate solution for the bounce.
    \item In the limit $M \rightarrow \infty$, the false vacuum decay rate becomes just like in the local theory (cf. Eq.~\eqref{local_limit}), which is consistent in any nonlocal theory in weak coupling limits.
    \item We observed that the nonlocal effects, implemented via a string-inspired nonlocal kinetic term for the scalar field smear the width of the bubble wall, and have no effect far away from it (cf. Figs.~\ref{plot_nloc_soliton_0_sigma} and \ref{plot_asympt_nloc_soliton_sigma}).
    \item We notice that when the scale of nonlocality is close to the mass scale of the potential $\mu$, the false vacuum decay rate is exponentially suppressed (cf. Eq.~\eqref{Q_result_nl} and Fig.~\ref{plot_Q}).
    \item For any given nonlocal scale, we showed that the thin-wall approximation breaks down when $M \sim \mu$, because of the smearing effect of nonlocality on the bubble wall (cf. Fig.~\ref{plot_Q}).
    \item We computed the bubble wall energy density and found that it is independent of the microscopic scales $\mu$ and $M$, as long as the thin-wall approximation is valid (cf. Eq.~\eqref{wall_energy}).
    \item Our bounce solution should be also valid for higher-derivative scalar field theories (Lee-Wick theories \cite{Grinstein:2007mp, Carone:2008iw}), with derivatives of finite largest order $N \gg 1$. Indeed, it was shown in Ref.~\cite{Boos:2021chb,Boos:2021jih,Boos:2021lsj} that such theory is asymptotically nonlocal when $N \rightarrow \infty$, such that it is equivalent to our nonlocal model. For $N \sim 1$, a specific analysis is needed that is beyond the scope of the present draft.
    \item Recently, a new recipe was proposed to build (from a ``mother'' local QFT) a UV-complete weakly nonlocal QFT that one can apply to gravity coupled to gauge and matter fields \cite{Modesto:2021ief}. Such construction is ghost-free in the nonlocal regime even in presence of a shift of vacuum, like in the Higgs mechanism \cite{Modesto:2021okr}, without relying on UV-completion in string theory. Other particularities are: (i) the tree-level scattering amplitudes are identical to the ones of the mother local QFT \cite{Modesto:2021soh}, with the same causal properties \cite{Giaccari:2018nzr}; (ii) any solution of the classical field equations of the mother local theory is also solution of the nonlocal one. Such properties do not hold in the class of string-inspired nonlocal scalar theory \cite{Buoninfante:2018mre} studied in our article. An important consequence is that, for these new theories, our semi-classical analysis will give the same results as the Coleman analysis in the local theory \cite{Coleman:1977py}, such that one needs to compute the quantum corrections to the bounce \cite{Callan:1977pt} in order to quantify the effect of weak nonlocality on the vacuum decay rate.
    \end{itemize}

Therefore, from our comparison between approximate analytical solutions and numerical solutions with the boundary conditions can help to understand 
the diligence of the diffusion method that have been employed. Particularly, we have found that the kink-type profiles used to describe non-perturbative transitions between non-degenerate vacua of the theory. Now since string-inspired nonlocal theories provide 
simplified playgrounds for the study of nonlocality in inflationary models, the search for background solutions could be of practical importance for alternative cosmological scenarios and also extend to strongly-coupled regimes. We envisage that our results will pave the way to understand the finite temperature effects on nonlocal solitons and understanding of the critical temperatures of such systems, with further applications in phase transitions and early universe cosmology including predictions of GW (from inflation, strong first-order phase transitions, etc.) in current and future GW detectors and recently searched by LIGO \cite{LIGOScientific:2019vic,KAGRA:2021mth,Romero:2021kby}. However, such study is beyond the scope of the present manuscript and will be taken up in future publication.

\acknowledgments

Authors acknowledge Jens Boos, Gianluca Calcagni, Anupam Mazumdar, Leonardo Modesto, Marco Frasca and Masahide Yamaguchi for useful comments.

\bibliographystyle{JHEP}

\providecommand{\href}[2]{#2}\begingroup\raggedright\endgroup

\end{document}